\documentclass[9pt,a4paper,twocolumn]{scrartcl}

\usepackage{lmodern}
\usepackage[utf8]{inputenc}
\usepackage[TS1,T1]{fontenc}
\usepackage[usenames,dvipsnames,svgnames,table]{xcolor}
\usepackage{url}
\usepackage{graphicx}
\DeclareGraphicsExtensions{.eps}
\usepackage{latexsym}
\usepackage{xspace}
\usepackage{enumitem}
\setenumerate{noitemsep,topsep=0pt,parsep=0pt,partopsep=0pt}
\usepackage[dvips]{hyperref}
\hypersetup{
bookmarks=true,
bookmarksnumbered = true,
hypertexnames=false,
pdfauthor={Serge Abiteboul, Émilien Antoine, Gerome Miklau, Julia
Stoyanovich, Jules Testard},
pdftitle={webdam system}, pdfsubject={webdamlog implementation},
pdfkeywords={webdamlog, datalog, distributed, declarative, provenance, graph},
pdfcreator={latex with hyperref package}}
\usepackage[hyperpageref]{backref}
\usepackage{amsmath}
\usepackage{balance}

\newcommand*{\eat}[1]{}
\newcommand{\dollar}{{\$}}

\newcommand*{\wdl}{WebdamLog\xspace}

\newcommand*{\bud}{Bud\xspace}
\newcommand*{\server}{\emph{Webdam cloud}\xspace}
\newcommand*{\soft}{Wepic\xspace}
\newcommand*{\confe}{\textsc{\textsf{sigmod}}\xspace}
\newcommand*{\confgroup}{\textsc{\textsf{SigmodFB}}\xspace}
\newcommand*{\usera}[1]{Émilien#1}
\newcommand*{\userb}[1]{Jules#1}
\newcommand*{\userj}[1]{Julia#1}

\newcommand*{\mypic}{\emph{My~picture}s\xspace}
\newcommand*{\otpic}{\emph{Attendee~pictures}\xspace}
\newcommand*{\querytab}{\emph{Query~tab}\xspace}

\newcommand*{\peer}[1]{\ensuremath{\mathsf{#1}}}
\newcommand*{\relName}[1]{\ensuremath{\mathsf{#1}}}
\newcommand*{\relduo}[2]{\relName{#1}\textbf{@}\peer{#2}}
\newcommand*{\reltri}[3]{\relName{#1}\textbf{@}\peer{#2}(\ensuremath{\mathsf{#3}})}
\newcommand*{\var}[1]{\ensuremath{\mathsf{\$}\mathsf{#1}}}

\newcommand{\rel}[3]{}

\newcommand{\serge}[1]{[\textcolor{red}{{\bf Serge: }#1}]}

\definecolor{julcol}{rgb}{0.71,0.21,0.19}

\definecolor{emicol}{rgb}{0.09,0.51,0.19}

\colorlet{comcol}{black!60}

\newenvironment{nstabbing}
{\setlength{\topsep}{-\parskip}%
\setlength{\partopsep}{0pt}%
\tabbing}
{\endtabbing}

\begin{document}
\balance

\title{Rule-Based Application Development using Webdamlog
\footnote{
This work has been partially funded by the European Research
Council under the European Community's Seventh Framework Program
(FP7/2007-2013); ERC grant Webdam, agreement
226513. \url{http://webdam.inria.fr/}}
}

\author{
% 1st. author
Serge Abiteboul\\
Inria Saclay \& ENS Cachan\\
first.last@inria.fr
\and
% 2nd. author
Émilien Antoine\\
Inria Saclay \& ENS Cachan\\
first.last@inria.fr
\and
% 3rd. author
Gerome Miklau\\
Inria Saclay \& UMass Amherst\\
miklau@cs.umass.edu
\and
% 4th. author
Julia Stoyanovich\\
Drexel University \& Skoltech\\
stoyanovich@drexel.edu
\and
% 5th. author
Jules Testard\\
Inria Saclay \& McGill U.\\
Jules.Testard@mail.mcgill.ca
}

%\date{\today}

\maketitle

\begin{abstract}
    We present the \wdl system for managing distributed data on the Web
    in a peer-to-peer manner.  We demonstrate the main features of the
    system through an application called \soft for sharing pictures
    between attendees of the \confe conference. Using \soft, the
    attendees will be able to share, download, rate and annotate
    pictures in a highly decentralized manner.  We show how \wdl handles
    heterogeneity of the devices and services used to share data in such
    a Web setting. We exhibit the simple rules that define the \soft
    application and show how to easily modify the \soft application.
\end{abstract}

% short version not in ACM  format
% ~\linebreak
% \noindent
% Category: H.2.4 [Systems]: Rule-based databases
% \noindent
% Keywords: datalog, peer to peer, declarative
% ACM format
% \category{H.2.4}{Systems}{Rule-based databases}
% \keywords{datalog, peer to peer, declarative}

%!TEX root = 12DemoWebdam.tex

\section{Introduction}
\label{sec:introduction}

Information management on the Internet relies on a wide variety of
systems, each specialized for a particular task. A user's
data and favorite applications are often distributed, making the
management of personal data and knowledge (i.e., programs) a major challenge.  Consider {\em Joe}, a typical Web user who has a
blog on Wordpress.com, a Facebook account, a Dropbox account, and
also stores data on his smartphone and laptop.  {\em
Joe} is a movie fan and he wants to post on his blog a review of
the last movie he watched. He also wishes to advertise his review to his Facebook friends and to include a link to his Dropbox folder where the movie has been uploaded. This is a cumbersome task to carry out manually, yet automating it, for example by writing a script, is far
beyond the skills of most Web users.

Some systems attempt to provide integrated services to support such
needs.  For instance, Facebook provides a wrapper service to integrate
Dropbox accounts and blogs.  However, such services are often limited
in the functionality they support.  Also, by delegating such services
to systems like Facebook, a user needs to trust more and more of his
information to one particular system.  Our goal is to enable the user
to easily specify distributed data management tasks \emph{in place}
(i.e., without centralizing his data to a single provider), while
allowing him to keep full control over his own
data~\cite{ICDE12:distributedKnowledgeBase}.  Our system is not a
replacement for Facebook, or any centralized system, but it allows
users to launch their customized peers on their machines with their
own personal data, and to collaborate using Web services.

This demonstration presents the \soft application, a distributed
picture manager. The \soft application is specified using simple rules written in a declarative language called~\wdl~{\cite{lang:webdamlog}} and it runs on the \wdl system,
demonstrated here for the first time.  \wdl is a rule-based datalog-style
language that emphasizes \emph{cooperation} between
\emph{auto\-no\-mous} peers by allowing both data, and
programs that compute on data, to be easily exchanged. The declarative approach alleviates the conceptual complexity on the user while, at the same time, allowing for powerful performance optimizations on the part of the system.

A central issue in such a setting is the ease with which a casual user
can write \wdl rules.  We conducted a user study, showing that users are
able to both understand and write simple \wdl programs with a minimal
amount of training~\cite{wlengine}. This demonstration further argues for
the simplicity of using \wdl for designing applications that handle
personal data.

\confe attendees will use \soft to share, download, rate and annotate
pictures taken at the conference.  Attendees will use the application
via a Web GUI or launch their own \soft peer.  They will first inspect
the basic \wdl rules of the provided application and will then be
invited to customize the application by modifying or adding rules.

\section{Language and System}
\label{sec:system}

\paragraph*{\wdl language, in brief} \label{sec:language-wdl} \wdl
extends datalog in a number of ways, supporting updates, distribution,
negation, and a novel feature called delegation.

\textbf{Facts.} A \emph{fact} is an expression of the form $m@p(a_1,
..., a_n)$, where $a_1, ..., a_n$ are data values and $m@p$ is a
relation described by relation name $m$ and peer name $p$.
Variables start with the symbol \dollar, e.g. \dollar$x$,
and data values are quoted. An example of a fact is: \\
\hspace*{0.5cm}
\reltri{pictures}{\confe}{\text{32},\text{``sea.jpg''},\text{``\usera{}''},\text{ 100\dots},~\dots}\\
where pictures@{\confe} is a relation managed at the \peer{\confe} peer
to represent a collection of \relName{pictures}. The picture name is
``sea.jpg'', its content is in binary ``100\dots'', and the
remainder of the fact consists of meta-data about the picture.

\textbf{Rules.}
A \emph{term} is a constant or a variable.
A \emph{rule} in a peer $p$ is an expression of the
form:\\
\hspace*{0.5cm}\reltri{\var{R}}{\var{P}}{\var{U}} ~:-~
\reltri{\var{R_1}}{\var{P_1}}{\var{U_1}} , \dots ,
\reltri{\var{R_n}}{\var{P_n}}{\var{U_n}}\\
where \var{R} and \var{R_i} are
relation terms, \var{P} and \var{P_i} are peer terms, \var{U} and \var{U_i}
are vectors of terms.

\textbf{Distribution/Delegation.} \wdl supports distribution, allowing
atoms in the body of a rule to refer to relations on remote peers
(using variables denoting relations or peers).  The evaluation of such
a rule therefore requires the knowledge and processing of possibly
more than one peer.  In addition, delegation allows a peer to install
a rule at a remote peer, where the rule itself may refer to local or
remote relations.  Rule bodies in WebdamLog are evaluated from left to
right. The order matters, unlike in datalog where the order of
atoms in a rule body is not relevant.  Although negation is supported
by the language, it is not yet implemented in the \wdl
system.\linebreak

%\serge{consistency: datalog no capital; Web with capital. I chose majority vote :)}

\paragraph*{\wdl peers, in brief} \label{sec:wdl-peer} Each peer runs a
\wdl program, i.e., a set of rules, using a \wdl engine.  The evaluation
of datalog has been studied for decades but a renewed interest in
datalog \cite{activexml,hellerstein-pods, Cylog} has led to the recent
development of the \bud~\cite{bloom} datalog engine, which we use as
part of our implementation.  The \bud system supports efficient
communication between peers and updates of base facts.\linebreak

A computation stage of the \wdl engine is broken down into three
steps. First, the peer loads the inputs received from the remote peers
since the previous stage.  Second, the peer runs a fixpoint
computation of its program.  Third, the peer sends facts (updates) and
rules (delegations) to other peers.  A main novelty is the possibility
for \wdl rules to have variables as relation and peer names.  Another
main novelty is delegation, which leads to installing, at run-time
rules, at other peers. As a result, WebdamLog programs may be very
dynamic, peers may discover new peers and new relations and they may
acquire new rules as the result of delegations.

The \wdl system follows prior work on a P2P
system for sharing data called
WebdamExchange~\cite{webdb11:webdamexchange,webdamexchange:demo}. The present focus is very different (declarative specification and delegation in \wdl vs. security in WebdamExchange), so little of the previous system could be re-used.

\paragraph*{Wrappers} \label{sec:wrapper} The \wdl system has been
designed for integration with the personal data of regular Web users
who already use popular Web systems such as Facebook, Picasa or
Dropbox.  The \wdl system supports the management of data in these
systems by using wrappers. A wrapper to some existing system $X$
provides software that exports to \wdl one or more relations
corresponding to the data in $X$, as well as rules to access/update
this data.

For example, in the demonstration we will use a Facebook wrapper. For
a given Facebook user \usera{}, our wrapper will simulate
a peer $\mathsf{\usera{FB}}$ with two relations: \\
\hspace*{0.5cm}\reltri{friends}{\usera{FB}}{\var{userID},\var{friendName}}
\\
\hspace*{0.5cm}\reltri{pictures}{\usera{FB}}{\var{picID},\var{owner},\var{URL}}\\ These
two relations provide an abstract view of \usera{}'s Facebook data
relevant to this particular application, and can then be used in \wdl
rules.

\paragraph*{Access control} \label{sec:access-control} Since many \wdl applications will manage personal or social data, access to sensitive information must be carefully controlled.  Access control in \wdl is particularly challenging because of the distributed nature of computation and the ability of peers to delegate rules to other peers.  We describe in the next section the features included in the demonstration for controlling the delegation of rules, in which a peer is asked to explicitly accept delegations initiated by other peers.

A complete access control model for \wdl is under active investigation and is not included in the demonstration. In that model, access to stored or derived relations is controlled by a novel combination of both discretionary methods (in which users have
the power to grant rights to data they own) and mandatory methods (in
which access rights are derived according to system-wide conventions).
Users directly specify the accessibility of stored relations that they
own.  For derived relations (i.e. views), a user may rely on a default
access control policy that is derived automatically from the
provenance of the base relations. Alternatively, a user may override
this policy in order to grant access to views, effectively
``declassifying'' some data.  This flexible model subsumes the
view-based access control of the standard SQL authorization model.

%Based on the specification of the access control policy, the data management rules written by the user are \emph{compiled} into internal rules that enforce that policy.

%%% Local Variables:
%%% mode: latex
%%% TeX-master: "sigmod376-Abiteboul"
%%% End:

\section{The Wepic application}
\label{sec:soft-application}

To demonstrate the \wdl system, we introduce an application called
\soft, a conference picture manager for the \confe conference. This
application allows conference attendees to share their pictures and
to rate, annotate and download the pictures of others.

\begin{figure}[htb]
    \centering
    \includegraphics[width=\linewidth]{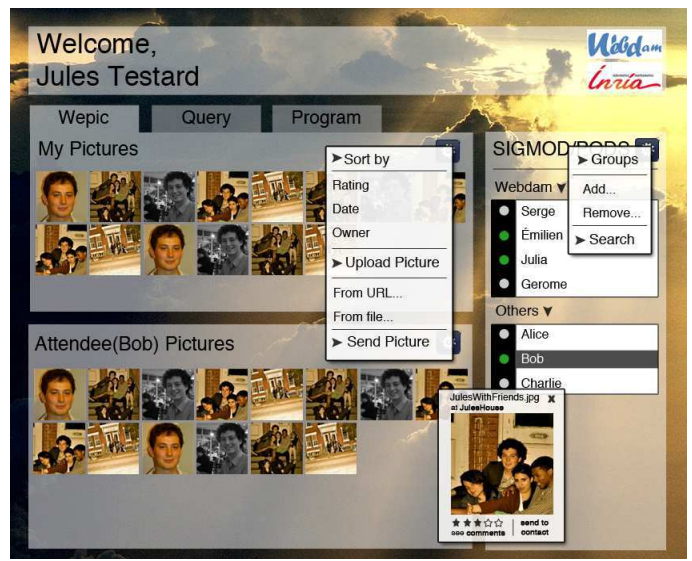}
    \caption{A screenshot of the \soft user interface.}
    \label{fig:screenshot_user_interface}
\end{figure}

\soft consists of a small set of rules that we discuss further. In
addition, the application uses two standard wrappers we implemented,
one for Facebook, and one for email communications. The \wdl system
provides a graphical interface, which has been customized to provide a
user interface for \soft. It is shown in
Figure~\ref{fig:screenshot_user_interface}.

A user connected to a \soft peer can:
\begin{enumerate}
    \item Upload a picture from a file or a URL;
    \item \label{item:seeviewattendee} View pictures
      provided by a particular attendee;
    \item Transfer pictures:
    \begin{enumerate}
        \item \label{item:sendpictoattendee} send them by email to
          the \confe group on Facebook, or to another \soft peer,
      \item get pictures from another \soft peer or from the \confe
        group on Facebook;
    \end{enumerate}
    \item \label{item:annotatepic} Annotate pictures with ratings,
    comments or name tags (names of attendees appearing in the picture);
    \item Select and rank photos based on their annotations.
\end{enumerate}

\eat{
The \soft program processed by the \wdl peer consists of a small
set of rules. To illustrate, we exhibit some rules used at peer
\userb{}.}

We now illustrate how some of these units of functionality are
implemented with \wdl rules.  To view pictures uploaded by a
particular \confe attendee,\eat{ For
  Functionality~\ref{item:seeviewattendee},} we use a relation
\relName{selectedAttendees} that contains one fact for each currently
highlighted attendee (see right-hand side column in
Figure~\ref{fig:screenshot_user_interface}). We also use a derived
relation \relName{pictures}, which is the view of all the pictures of
a particular attendee. To obtain the pictures of all selected
attendees, we use the rule:
\begin{nstabbing}
    \hspace*{5em}\= \kill
    \hspace*{0.5cm} \reltri{attendeePictures}{\userb{}}{\var{id},\var{name},\var{owner},\var{data}} :-\\
    \>\reltri{selectedAttendee}{\userb{}}{\textcolor{Red}{\var{attendee}}},\\
    \>\reltri{pictures}{\textcolor{Red}{\var{attendee}}}{{\var{id}},\var{name},\var{owner},\var{data}}
\end{nstabbing}
Note that this rule uses delegation, a feature novel to \wdl, to
retrieve the contents of relation \relName{pictures} of each attendee.
The result of executing this rule is shown in the \otpic
frame at the bottom of Figure~\ref{fig:screenshot_user_interface}.

To transfer pictures between peers, we assume that each attendee
specifies some preferred communication protocols in relation
\relName{communicate}, stating, e.g., whether he prefers to receive
pictures by email, by posting on Facebook, or directly in his Wepic
peer.  The following rule is executed when Jules sends some pictures
to some attendees: \eat{ If the user of the peer changes her habits,
  the pictures can be sent via the right protocol:}
\begin{nstabbing}
    \hspace{5em}\= \kill
    \hspace*{0.5cm}\reltri{\textcolor{Blue}{\var{protocol}}}{\textcolor{Red}{\var{attendee}}}{\textcolor{Red}{\var{attendee}},\var{name},\var{id},\var{owner}} :-\\
    \>\reltri{selectedAttendee}{\userb{}}{\textcolor{Red}{\var{attendee}}},\\
    \>\reltri{communicate}{\textcolor{Red}{\var{attendee}}}{\textcolor{Blue}{\var{protocol}}},\\
    \>\reltri{selectedPictures}{\userb{}}{\var{name},\var{id},\var{owner}}
\end{nstabbing}

Rules of this kind, and other rules implementing the basic
functionality of \soft, are available for inspection and customization
through the user interface, see Figure~\ref{fig:screenshotprogram}.

\paragraph*{Delegation and access control}

\eat{The \soft application contains sensitive information and access needs
to be carefully controlled.  As previously mentioned, rule delegation
in \wdl presents additional challenges. }

As noted in Section~\ref{sec:system}, by using delegation a user may write
a rule and ask another peer to process it remotely.  Consider
again the previous rule:
\begin{nstabbing}
    \hspace*{5em}\= \kill
    \hspace*{0.5cm} \reltri{attendeePictures}{\userb{}}{\var{id},\var{name},\var{owner},\var{data}} :-\\
    \>\reltri{selectedAttendee}{\userb{}}{\textcolor{Red}{\var{attendee}}},\\
    \>\reltri{pictures}{\textcolor{Red}{\var{attendee}}}{\var{id},\var{name},\var{owner},\var{data}}
\end{nstabbing}
Suppose we have the facts:\\
\hspace*{0.5cm}\reltri{selectedAttendee}{\userb{}}{\textcolor{Red}{\text{``\usera{}''}}}\\
The evaluation of the rule leads to delegating the following rule to
\usera{}:
\begin{nstabbing}
    \hspace*{5em}\= \kill
    \hspace*{0.5cm} \reltri{attendeePictures}{\userb{}}{\var{id},\var{name},\var{owner},\var{data}} :-\\
    \>\reltri{pictures}{\textcolor{Red}{Émilien}}{{\var{id}},\var{name},\var{owner},\var{data}}
\end{nstabbing}
to send all the facts in his relation \text{pictures} to \userb{}.  This
is a simple case of delegation, which can be controlled by {\em
inferring access} from the specifications described above.  However
delegated rules can be more complex, and general methods for effectively
controlling delegation are a topic of continued investigation.  The
demonstration of \soft will provide a simplified model for control of
delegation, in which each delegation sent by an untrusted peer will be
pending in a queue until the user explicitly accepts it via the Web
interface. A notification of a pending delegation can be seen at the top
of Figure~\ref{fig:screenshotprogram}, where \userj{} is sending a rule
to \userb{}.  By default, all peers except the \confe peer will be
considered untrusted.\eat{ so the \confe peer will not accept
delegations.}

%%% Local Variables:
%%% mode: latex
%%% TeX-master: "sigmod376-Abiteboul"
%%% End:

%!TEX root = 12DemoWebdam.tex

\section{Demonstration Scenarios}
\label{sec:scenario}

We now describe the general proceedings of the demonstration. The goal
will be to share pictures taken during the \confe conference. \usera{}
and \userb{} are attendees of the conference.  They have used \soft to
install locally on their laptops a collection of pictures.  They
demonstrate how to use \soft with the native functionalities described
in Section~\ref{sec:soft-application} and how to customize the
application.  They will also allow a user at the conference to run his
own \soft peer to explore the system. This scenario will demonstrate
the various aspects of \wdl, notably distribution, delegation and
control of delegation.

\begin{figure}[htb]
    \centering
    \includegraphics[width=0.98\linewidth]{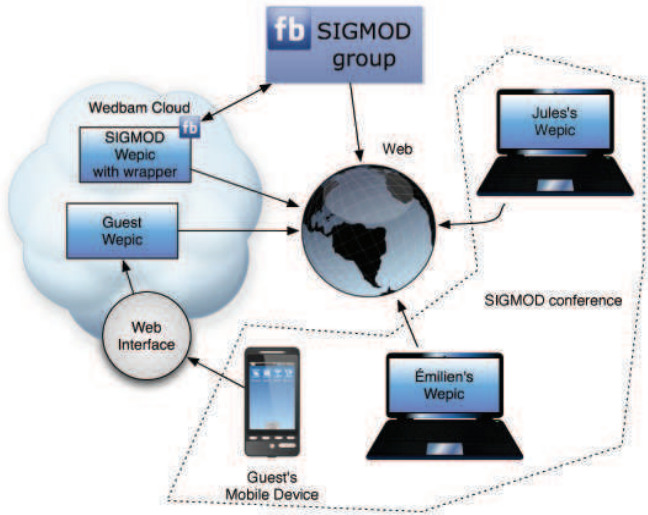}
    \caption{The distribution of peers in the network.}
    \label{fig:peernetwork}
\end{figure}

\eat{The distribution of the peers mentioned in the following
  description is shown in Figure~\ref{fig:peernetwork}.}

\begin{figure*}[tb]
    \centering
    \includegraphics[width=10cm]{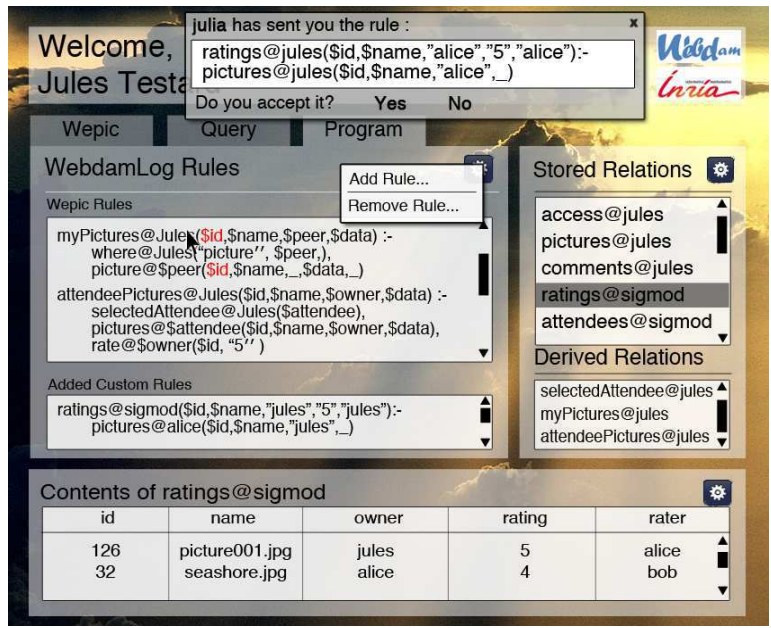}
    \caption{The interface to a \wdl program running \soft.}
    \label{fig:screenshotprogram}
\end{figure*}

\paragraph*{Setup} \label{sec:setup} In the beginning of the demo,
three peers are established: one on each of the laptops of
\usera{} and \userb{}, connected via a local network, and a third, the
\confe peer, hosted on \server. To simplify the presentation, we will
assume that \usera{} and \userb{} organize their data and behave
similarly.  They both store their photos in
\relduo{pictures}{\usera} and \relduo{pictures}{\userb} on their
respective \soft peers.  Both have Facebook accounts and are members
of the \confgroup group, the official Facebook group of the
conference.  Finally, both users are subscribed to the \confe peer,
which stores the list of registered \soft users.

The \soft peers \usera{}, \userb{} and \confe are depicted in
Figure~\ref{fig:peernetwork}.  We will start the demonstration by
quickly going over the setup, and will then ask the user to interact
with \soft according to the following scenarios.

\eat{ the basis for explaining functionalities of the system. During
  the demonstration, after rapidly explaining the setting, we will go
  over the following scenarios.}

\paragraph*{Interaction via Facebook}
\label{sec:inter-via-faceb}
To illustrate the interaction between a \soft peer and other Web
services, we use a Facebook wrapper. For instance, the following rule
is used by the \confe peer to automatically publish, on the Facebook
group of \confe, the pictures belonging to \confe attendees who have
authorized this action:
\begin{nstabbing}
    \hspace{5em}\= \kill
    \hspace*{0.5cm}
    \reltri{pictures}{\confgroup}{\textcolor{Red}{\var{id}},\var{name},\textcolor{Blue}{\var{owner}},\var{data}} :-\\
    \>\reltri{pictures}{\confe}{\textcolor{Red}{\var{id}},\var{name},\textcolor{Blue}{\var{owner}},\var{data}},\\
    \>\reltri{authorized}{\textcolor{Blue}{\var{owner}}}{\text{``Facebook''},\textcolor{Red}{\var{id}},\textcolor{Blue}{\var{owner}}}
\end{nstabbing}

Conversely, the \confe peer will automatically retrieve the pictures
with their comments and tags from the Facebook group and publish them
to \confe peer. Note that the system thus allows any \soft user to see
or publish (via \soft) pictures in \confgroup even without having a
Facebook account.  A user only needs to appropriately populate
his \relName{authorized} relation to control Facebook publication.

We will explain the \wdl rules that implement these interactions to
audience members. We will then show that a photo uploaded by \usera{}
into his local relation \relduo{pictures}{\usera} is instantly
published to \relduo{pictures}{\confe}, and then propagated to
\relduo{pictures}{\confgroup}.

\eat{ We will then illustrate how easy it is to customize such rules.
  In particular, we will show how the program of the \confe peer can
  be changed to restrict the list of photos retrieved from Facebook,
  by filtering on tags.}

\paragraph*{Customizing rules}
\label{sec:add-rules}
The main advantage of a peer-to-peer system such as \wdl is the
ability to customize a peer's behavior.\eat{Since it would be
  difficult to install a \soft peer on the laptop or smartphone of an
  attendees, we propose an alternative solution. The attendee can
  connect to the Web interface on \server to launch their own
  dedicated peer on which they can upload their photos and modify
  their program, as we do on the laptop-based peers.  Such an
  interaction is shown in Figure~\ref{fig:screenshotprogram}.}
Therefore the most novel trait of \soft is that it lets the user
customize existing rules and add his own rules.  For example, a user
who is interested only in the pictures that have a rating of 5 would
customize the rule of the application as follows:
\begin{nstabbing}
    \hspace{5em}\= \kill
    \hspace*{0.5cm}
    \reltri{attendeePictures}{\userb{}}{\textcolor{Red}{\var{id}},\var{name},\textcolor{Blue}{\var{owner}},\var{data}} :-\\
    \>\reltri{selectedAttendee}{\userb{}}{\textcolor{Green}{\var{attendee}}},\\
    \>\reltri{pictures}{\textcolor{Green}}{\textcolor{Red}{\var{id}},\var{name},\textcolor{Blue}{\var{owner}},\var{data}},\\
    \>\reltri{rate}{\textcolor{Blue}{\var{owner}}}{\textcolor{Red}{\var{id}},\text{5}}
\end{nstabbing}

Redefining this rule will change the contents of the frame \otpic in
Figure~\ref{fig:screenshot_user_interface}, which we will demonstrate.
We will explain this rule and its effect to audience members, and will
then allow them to customize the rule further, retrieving, e.g., only
pictures that were taken by a certain \confe attendee, or in which only
certain attendees appear.\eat{ This customization is easy to implement
by incorporating picture meta-data.}

\paragraph*{Illustration of the control of delegation}
\label{sec:delegation-access}
To illustrate the control of delegation, \usera{} will
attempt to install a rule at \userb{}' peer.  We will demonstrate
that the system requires the approval of \userb{} before installing
the rule, and that the program of \userb{} is changed once the
approval is granted and the rule is installed.

% This is a scenario that few users will opt for, so I'm moving it
% down the list
\paragraph*{Interaction via the Web}
\label{sec:inter-with-smartph}
Finally, we will invite audience members to launch their own autonomous
\soft peers in the \server, and to interact with their peer through a
UI, using their smartphone or a tablet that we provide.  From that point
on, they will be able to use all features of \soft described in
Section~\ref{sec:soft-application}, e.g., upload their conference photos
to their own peer.  In addition, they will be able to use the \querytab
to launch one of the pre-defined queries, or to write their own \wdl
queries and rules with the assistance of the demo authors.

%%% Local Variables:
%%% mode: latex
%%% TeX-master: "12DemoWebdam"
%%% End:

\bibliographystyle{abbrv}
% \small{\bibliography{demobibli}}
\bibliography{demobibli}

\end{document}